\documentclass{article}

\usepackage{tikz}

\usepackage[a4paper, portrait, margin=1in]{geometry}

\usepackage[round]{natbib}
\usepackage[colorlinks=true, citecolor=red, linkcolor=blue, urlcolor=green]{hyperref}
\usepackage{amsmath}
\usepackage{amsfonts}
\usepackage{cases}
\usepackage[affil-it]{authblk}
\usepackage{pdflscape}

\usepackage{graphicx}
\graphicspath{{./images}}

\title{Predicting Financial Market Crises using Multilayer Network Analysis and LSTM-based Forecasting of Spillover Effects}

\author{Mahdi Kohan Sefidi}
\affil{University of Khatam}
\date{May. 2025}

\begin{document}
\maketitle
\begin{flushleft}

\end{flushleft}
\begin{abstract}
    Financial crises often occur without warning, yet markets leading up to these events display increasing volatility and complex interdependencies across multiple sectors. This study proposes a novel approach to predicting market crises by combining multilayer network analysis with Long Short-Term Memory (LSTM) models, using Granger causality to capture within-layer connections and Random Forest to model interlayer relationships. Specifically, we utilize Granger causality to model the temporal dependencies between market variables within individual layers, such as asset prices, trading values, and returns. To represent the interactions between different market variables across sectors, we apply Random Forest to model the interlayer connections, capturing the spillover effects between these features. The LSTM model is then trained to predict market instability and potential crises based on the dynamic features of the multilayer network. Our results demonstrate that this integrated approach, combining Granger causality, Random Forest, and LSTM, significantly enhances the accuracy of market crisis prediction, outperforming traditional forecasting models. This methodology provides a powerful tool for financial institutions and policymakers to better monitor systemic risks and take proactive measures to mitigate financial crises.
\end{abstract}
\section{Introduction}
The complexity inherent in financial markets has long posed significant challenges for researchers and policymakers seeking to understand and predict systemic risks. The 2008 financial crisis exposed the fragility of the global financial system, underscoring the necessity of advanced methods capable of identifying early warning signals of financial instability \cite{yellen2013interconnectedness}. Traditional forecasting models, often based on linear correlations or simplistic representations of market dynamics, proved inadequate in capturing the multifaceted interactions that underpinned the crisis. In particular, the interconnectedness of financial institutions and the transmission of shocks across sectors remain critical areas for investigation, as they hold the key to understanding how disturbances in one part of the financial system can propagate across the global economy \cite{Sandhu2016Ricci}.

Recent advances in network theory have highlighted the limitations of single-layer models in representing the true complexity of financial systems. Single-layer models, which typically focus on individual market variables such as asset prices or returns, fail to account for the interdependencies across different sectors or market dimensions. As noted by \cite{levy2015dynamical} and \cite{diebold2014network}, financial systems are best understood as multilayer networks, where distinct types of connections—such as price, return, and trading value—interact in complex, often nonlinear ways. This multilayer perspective offers a more accurate representation of the structural relationships in financial markets and provides a foundation for more sophisticated methods of systemic risk prediction.

The multilayer network approach, as proposed by \cite{kivela2014multilayer} and \cite{boccaletti2014structure}, provides a foundational framework for understanding the complexity of financial systems. These works emphasize that networks in which multiple types of interactions (such as price changes, trading values, and credit risk) are accounted for simultaneously lead to a deeper understanding of market behavior. Additionally, \cite{de2013mathematical} introduced a mathematical formulation of multilayer networks that helps quantify and model how layers interact, offering valuable insights into the dynamics of complex systems.

The challenges of modeling such interconnections are compounded by the nonlinear and dynamic nature of financial markets, which require methods capable of adapting to changing conditions over time. Traditional methods are limited in their ability to capture the nonlinear interactions that often characterize financial systems \cite{Papana2017Financial}. Linear models often fail to distinguish between different types of market connectivity, such as those occurring during extreme market movements (tail events) versus normal periods \cite{Chen2023Financial}. Nonlinear Granger causality and quantile-based methods (e.g., quantile Granger causality) capture more intricate dependencies, especially during periods of financial stress or extreme market returns \cite{Bonaccolto2019Estimation}.

In this context, machine learning techniques have emerged as promising tools for overcoming the limitations of traditional methods. The integration of Granger causality, Random Forest, and Long Short-Term Memory (LSTM) models offers a robust framework for capturing the temporal, interlayer, and nonlinear dependencies that define financial market dynamics. Granger causality is particularly effective in modeling the temporal interdependencies within individual layers of the financial system, such as asset prices or returns \cite{Bonaccolto2019Estimation}. Random Forest, on the other hand, excels at modeling the complex relationships between layers, identifying how shocks in one sector can affect others \cite{Biau2010Analysis}. LSTM models, with their capacity to learn long-term dependencies from sequential data, provide a powerful tool for forecasting future market instability based on historical patterns \cite{Pilla2025Forecasting}, \cite{Abbasimehr2020An}.

Unlike traditional forecasting models, which often rely on linear assumptions and fail to capture the complex, nonlinear interactions that define financial systems, our approach integrates the strengths of Granger causality, Random Forest, and Long Short-Term Memory (LSTM) models to create a more robust prediction framework. Granger causality is well-established for modeling temporal dependencies, specifically within individual layers of the financial system, such as asset prices, trading volumes, or returns. By incorporating Granger causality, we can identify the causal relationships between variables over time, providing valuable insights into the evolution of market conditions. However, while Granger causality excels in capturing temporal dependencies, it has limitations in modeling the nonlinear and interdependent nature of relationships between different market sectors.

To overcome this, we use Random Forest to model the complex interlayer relationships that arise between various market variables across sectors. Random Forest, a machine learning algorithm known for its ability to handle high-dimensional, nonlinear data, captures how shocks in one sector can ripple across other sectors, which is crucial for understanding systemic risk. This approach is particularly effective for identifying complex patterns and interactions that traditional models may overlook.

Finally, we integrate Long Short-Term Memory (LSTM) networks to address the dynamic and sequential nature of financial data. LSTMs are particularly suited for time-series forecasting as they can learn long-term dependencies from historical data. This ability allows our model to adapt to changing market conditions over time, which is critical in predicting market instability or crises. The combination of LSTM with Granger causality and Random Forest creates a comprehensive model that captures both the temporal, causal relationships within market layers, as well as the complex, nonlinear interactions between market sectors.

By incorporating both within-layer and interlayer interactions, the proposed model provides a more comprehensive understanding of financial markets and offers a significant improvement over traditional crisis prediction models. Drawing on insights from previous research (e.g., \cite{billio2012econometric}, \cite{wang2017extreme}), this study aims to advance our understanding of systemic risk in complex financial systems and contribute to the development of more effective tools for financial institutions and policymakers to mitigate the risks of future crises.

In summary, by combining these three techniques, our model overcomes the limitations of traditional methods. Granger causality provides a foundation for understanding temporal relationships, Random Forest captures the intricate interlayer dynamics, and LSTM adapts to changing conditions, offering a more accurate and holistic prediction of financial crises. The proposed methodology not only enhances the accuracy of financial crisis prediction but also provides a deeper understanding of the underlying structures that govern market behavior. In doing so, it offers a powerful new tool for monitoring the stability of the global financial system and addressing the challenges posed by the increasing complexity and interconnectedness of financial markets.

In the following sections, we first explore single-layer models in Section~\ref{sec:single-layer}, where Granger causality is applied to capture temporal dependencies within individual market layers, such as asset prices and trading volumes. Section~\ref{sec:multilayer-network} extends this by introducing a multilayer network approach, utilizing Random Forest to model the complex interlayer relationships between market sectors and their respective variables.

Finally, in Section~\ref{sec:lstm-result}, we analyze network indicators and use Long Short-Term Memory (LSTM) models to predict market crises by capturing long-term temporal dependencies in market data. This integrated approach combines the strengths of Granger causality, Random Forest, and LSTM to enhance crisis prediction accuracy and provide deeper insights into the dynamics of financial instability.

\section{Single Layer}\label{sec:single-layer}

In this section, we consider the financial market as composed of several distinct layers, each representing a specific market indicator. Specifically, the indicators of price, return, and trading values are each modeled as a separate single layer within the network framework.

For each layer, the nodes correspond securities that are members of the S\&P 500 index. The relationships (edges) between these nodes are defined based on Granger causality tests applied to the time series data of each indicator. More precisely, for each pair of securities, a Granger causality test is conducted to determine whether the past values of one security's indicator at time \( t-1 \) can predict the current value at time \( t \) of another security's indicator within the same layer.

Thus, the single-layer network captures the temporal dependencies and directional causal relationships between securities for each individual market variable, providing a foundational representation of the market structure before incorporating cross-layer interactions.

\subsection{Mathematical Model of Granger Causality}

Granger causality tests whether past values of one time series \cite{Shojaie2021Granger}, \( X_t \), can be used to predict future values of another time series, \( Y_t \). In this paper, we apply Granger causality to three layers: asset prices, returns, and trading values. Specifically, we test the temporal dependencies within each of these layers for each security in the S\&P 500. The general autoregressive model for Granger causality is:

\begin{equation}
Y_t = \alpha_0 + \sum_{i=1}^{p} \beta_i Y_{t-i} + \sum_{i=1}^{p} \gamma_i X_{t-i} + \epsilon_t
\end{equation}

where:
\begin{itemize}
    \item \( Y_t \) is the dependent variable at time \( t \), which represents the time series we are trying to predict (e.g., asset prices, returns and trading values),
    \item \( X_t \) is the independent variable at time \( t \), which represents the potential cause (e.g., same type of indicator with \( Y_t  \) but in the different security),
    \item \( \alpha_0 \) is the intercept term,
    \item \( \beta_i \) are the coefficients associated with the lagged values of the dependent variable \( Y_t \),
    \item \( \gamma_i \) are the coefficients associated with the lagged values of the independent variable \( X_t \),
    \item \( p \) is the number of lags in the model (i.e., the number of previous time periods included for prediction),
    \item \( \epsilon_t \) is the error term at time \( t \), assumed to be white noise.
\end{itemize}

To test the relationship between \( X_t \) and \( Y_t \), we perform a hypothesis test using the p-value. If the p-value of the Granger causality test is smaller than a predefined significance level, \( \theta \), we conclude that \( X_t \) Granger-causes \( Y_t \). The significance level \( \theta \) serves as a threshold for determining whether the relationship between two time series is statistically significant. In this paper, we consider a p-value smaller than \( \theta \) (e.g., 0.05) as an indication of a statistically significant relationship.

The null hypothesis for Granger causality is that \( X_t \) does not Granger-cause \( Y_t \), implying that the coefficients \( \gamma_1, \gamma_2, \dots, \gamma_p \) are all zero. To test this, we perform an F-test to determine whether the joint null hypothesis \( H_0: \gamma_1 = \gamma_2 = \dots = \gamma_p = 0 \) can be rejected. If the F-test rejects the null hypothesis, we conclude that \( X_t \) Granger-causes \( Y_t \), meaning that past values of \( X_t \) provide valuable information for predicting \( Y_t \).

\subsection{Network of One Layer}

In this study, we define a network of one layer, where each node represents a market variable, such as asset prices, returns, or trading volumes. The network is mathematically represented as:

\[
\mathcal{N}_1 = (V_1, E_1)
\]
where:
\begin{itemize}
    \item \( V_1 = \{v_1, v_2, \dots, v_n\} \) is the set of \( n \) nodes, each representing a market variable or security in the financial system.
    \item \( E_1 \subseteq V_1 \times V_1 \) is the set of edges, where each edge \( (v_i, v_j) \in E_1 \) represents a significant relationship between nodes \( v_i \) and \( v_j \). This relationship can be a causal or statistical connection, such as Granger causality, indicating the influence or interdependency between the variables.
\end{itemize}

The network \( \mathcal{N}_1 \) captures the direct dependencies between market variables within a single layer, focusing on pairwise relationships. Each edge \( (v_i, v_j) \) reflects the directional influence of one variable on another, for instance, the predictive relationship between asset prices and returns. This single-layer network does not account for cross-layer interactions and is limited to the relationships within one market dimension (e.g., asset prices). By constructing this network, we can study how individual market variables interact with each other, laying the foundation for more complex multilayer network models that include interlayer dependencies.

\subsection{Interconnection}
The interconnection of each security in the network can be defined as the degree of each node, which represents the number of direct relationships (edges) it has with other securities. In the sparse matrix \( R \), the degree of node \( v_i \) is given by the sum of the values in its corresponding row, which indicates the number of securities that Granger-cause \( v_i \), or are Granger-caused by \( v_i \). Mathematically, the degree \( \deg(v_i) \) of node \( v_i \) is defined as:

\[
\deg(v_i) = \sum_{j=1}^{n} R_{ij}
\]

where the entries of the sparse matrix \( R \) are defined by the conditional function:

\begin{equation}
R_{ij} =
\begin{cases}
1 & \text{if the p-value } p_{ij} \leq \theta, \\
0 & \text{if the p-value } p_{ij} > \theta,
\end{cases}
\end{equation}

with \( p_{ij} \) representing the p-value of the Granger causality test between securities \( v_i \) and \( v_j \), and \( \theta \) denoting the significance threshold.

The degree \( \deg(v_i) \) quantifies the interconnection of security \( v_i \) with other securities in the network. A higher degree means that the security has more significant relationships (causal or statistical) with other securities in the system, indicating stronger interconnection. Conversely, a lower degree implies fewer significant relationships and weaker interconnection. Thus, the degree of each node provides a measure of the connectivity or influence of each security in the network based on its Granger-causal relationships with other securities in the S\&P 500.

\section{Multilayer Network}\label{sec:multilayer-network}

In the multilayer network framework, the financial market is modeled as interconnected layers, each representing a distinct market indicator such as Price, Return, and Trading Value. Unlike single-layer networks that capture relationships within one type of variable, multilayer networks explicitly model the directional causal relationships between these different layers to reflect potential spillover effects. Specifically, we focus on three key interlayer relations that are meaningful for understanding how information and shocks propagate through the market: \(\text{Price} \rightarrow \text{Return}\), \(\text{Trading Value} \rightarrow \text{Price}\), and \(\text{Trading Value} \rightarrow \text{Return}\). These relations are identified and quantified using the Random Forest algorithm, which captures complex nonlinear dependencies and interactions across layers. By applying Random Forest to the lagged data of nodes across different layers, we can robustly estimate the strength and significance of these causal spillover relationships, providing a comprehensive picture of cross-layer market dynamics.

\subsection{Multilayer Network Structure}

We define the multilayer network \( \mathcal{N}_M \) as a set representing the interlayer relationships between nodes across different layers. Specifically, the network is:

\[
\mathcal{N}_M = (V, E)
\]

where:
\begin{itemize}
    \item \( V = \{v_1, v_2, \dots, v_n\} \) is the set of nodes, each representing a market variable (e.g., asset prices, returns, trading values).
    \item \( E \subseteq V_1 \times V_2 \cup V_2 \times V_3 \cup \dots \cup V_{L-1} \times V_L \) is the set of edges representing the interlayer relationships, where each edge connects a node in one layer to a node in a different layer. For example, an edge between \( v_i^{(1)} \in V_1 \) (Layer 1) and \( v_j^{(2)} \in V_2 \) (Layer 2) indicates the relationship between the asset price in Layer 1 and the return in Layer 2.
\end{itemize}

Thus, \( \mathcal{N}_M \) only includes edges between nodes across different layers, ensuring that the model captures the interactions between variables from different market dimensions (e.g., asset prices and returns) but does not account for relationships within the same layer.

To model the relationships between nodes in different layers, we apply Random Forest, a powerful machine learning technique that can capture nonlinear dependencies. Specifically, Random Forest is used to predict the relationship between nodes of one layer based on the nodes of another layer, while incorporating a time lag. This time lag allows us to model the spillover effect between market dimensions.

The spillover effect refers to the influence that variables in one layer, such as asset prices in Layer 1, can have on variables in another layer, such as returns in Layer 2, over time. This effect captures how shocks or changes in one market dimension can propagate to another. For instance, a large change in asset prices may spill over into the returns layer, affecting the financial performance of other securities. Modeling this spillover is crucial for understanding systemic risk and the interdependencies within the financial system.

Let the feature set \( \mathcal{X}_l \) represent the features (market variables) in layer \( l \), and let \( \mathcal{Y}_l \) represent the target variable(s) for that layer (e.g., returns or asset prices). We apply Random Forest to predict \( \mathcal{Y}_l \) based on \( \mathcal{X}_l \), while incorporating the time lag between the nodes of different layers. This allows us to capture how past values of variables in one layer influence future values in another layer.

For example, the relationship between asset prices in Layer 1 (\( v_i^{(1)} \)) and returns in Layer 2 (\( v_j^{(2)} \)) is modeled by applying Random Forest with a time lag of \( p \), where past asset prices in Layer 1 are used to predict future returns in Layer 2. This time lag captures the spillover effect, which reflects how movements in asset prices affect returns with a delay.

In the Random Forest model, the relationship between the nodes is quantified by the importance scores, which measure the strength of the spillover. If a node from one layer significantly influences a node in another layer, it will have a high importance score. These importance scores help define the edges in the multilayer network.

\subsection{Relation Model with Feature Reduction}

After applying Random Forest, we perform feature reduction to remove irrelevant or redundant features based on the importance scores derived from the model. This process helps concentrate on the most influential variables and reduces the overall model complexity.

Following feature reduction, we introduce a threshold \( \zeta \) to decide which interlayer relationships (edges) should be included in the multilayer network. The importance score for each pair of nodes, obtained from the Random Forest output, is compared against this threshold. If the importance score exceeds \( \zeta \), the edge between those nodes is retained; otherwise, it is discarded. Formally, the entries \( S_{ij} \) of the sparse adjacency matrix representing interlayer relationships are assigned according to the condition:

\[
S_{ij} = 
\begin{cases}
1 & \text{if } \text{importance}_{ij} > \zeta, \\
0 & \text{otherwise},
\end{cases}
\]

where \(\text{importance}_{ij}\) denotes the importance score of the relationship between nodes \( v_i^{(l)} \) and \( v_j^{(l')} \), and \(\zeta\) is the threshold for retaining significant edges.

This thresholding ensures that only significant interlayer connections remain, thereby improving the network’s clarity and interpretability. The resulting sparse matrix \( S \) encodes these relationships, with 1 indicating a significant link and 0 indicating no link between nodes of different layers.

In addition to modeling intralayer dynamics within each layer, we emphasize interlayer relationships to capture how variables in one layer influence those in another over time. These interlayer dependencies are modeled using Random Forest with an incorporated time lag between layers. This lag allows us to capture the temporal effect where past values of one market dimension (e.g., asset prices) affect future values in another dimension (e.g., returns).

By employing Random Forest with this time lag, we effectively capture both temporal dependencies and cross-layer interactions, providing a comprehensive representation of the complex interrelationships in the financial system.

\section{Data Analysis and Prediction Results}\label{sec:lstm-result}

In this section, we present the results of our analysis of the relationships between securities in the S\&P 500. We focus on modeling the interactions between securities using Granger causality, and we visualize these relationships through heatmaps. The results offer insights into the temporal evolution of these relationships and highlight how market variables such as asset prices, trading values, and returns influence one another over time. The findings from this analysis are crucial for understanding the dynamics of financial networks and assessing the spillover effects between market layers.

\subsection{Dataset and Parameters}

The dataset used in this study consists of financial data for the S\&P 500 index and 140 securities that are members of the S\&P 500. The data spans from January 1, 2000, to April 2025. For each security, we calculate three key financial indicators: \textit{price}, \textit{trading value}, and \textit{return}. These indicators are computed over a sliding window approach with a window size of 100 days, and the window steps forward by 30 days at each iteration.

For each indicator (price, trading value, and return), we calculate the \textit{mean} and \textit{minimum} values over the 100-day sliding window for each step. This methodology provides time series data for each of the three indicators over the entire dataset period. The time points \( t_i \) are defined based on the sliding window, where \( t_0 \) corresponds to the initial window (starting at day 1), and subsequent time points \( t_1 \), \( t_2 \), etc., correspond to the following windows with a 30-day step.

\subsection{Evolution of layers}

The relationships between the 140 securities are modeled using Granger causality over three time intervals: \( t_0 \) to \( t_1 \), \( t_1 \) to \( t_2 \), and \( t_2 \) to \( t_3 \). For each time interval, the p-values from the Granger causality tests are used to determine the strength of the relationships between each pair of securities across the three layers (price, trading value, and return). 

\begin{figure}[ht]
    \centering
    \includegraphics[width=\linewidth]{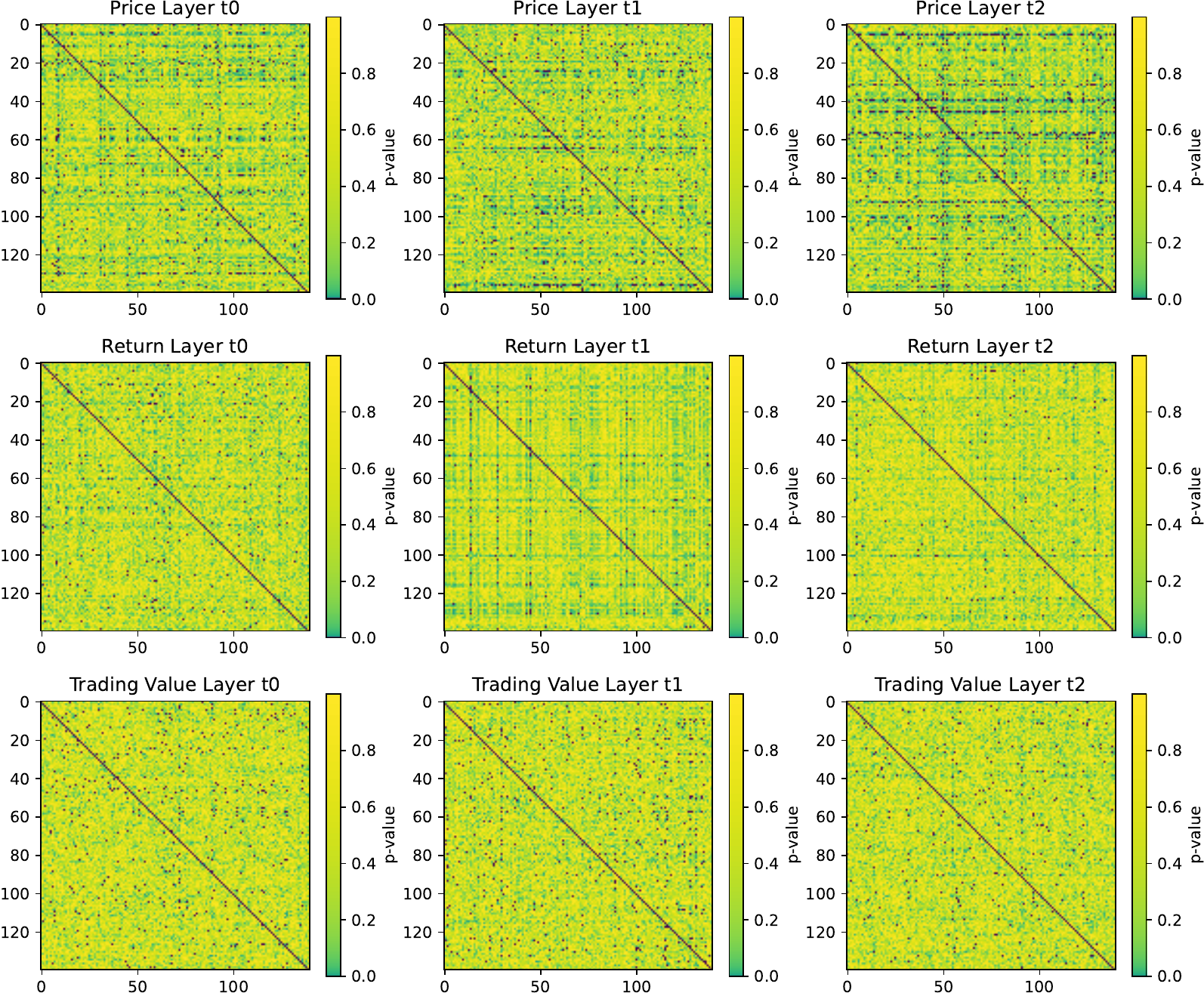}
    \caption{Heatmap of the interrelationship network based on p-values from Granger causality tests for the time intervals \( t_0 \) to \( t_2 \) for each indicator (price, trading value, and return). The color intensity represents the strength of the relationship, with darker shades indicating stronger relationships.}
    \label{fig:colormap-single-layer}
\end{figure}

A heatmap is generated to visualize the network based on the p-values for Granger causality tests, as shown in Figure \ref{fig:colormap-single-layer}. This heatmap represents the interrelationships between securities based on the significance of the Granger causality p-values over the three time intervals. Each cell in the heatmap corresponds to the p-value for the Granger causality test between a pair of securities, where darker colors indicate a stronger relationship (lower p-value) and lighter colors indicate weaker relationships (higher p-value).

The heatmap shown in Figure~\ref{fig:colormap-single-layer}, offers a visual representation of how the relationships between securities evolve over time. By observing the changes in color intensity across the three time intervals \( t_0 \) to \( t_2 \), we can identify which securities exhibit persistent relationships across different market dimensions (price, trading value, and return). Stronger relationships are represented by darker colors, while lighter colors signify weaker or insignificant relationships. The heatmap also helps assess the degree of spillover between layers, where stronger interlayer relationships are represented by darker cells. This enables a better understanding of how changes in one market dimension (such as asset prices) spill over into other dimensions (such as returns).

The graph in Figure~\ref{fig:colormap-single-layer} provides insights into the dynamic relationships between securities. Notably, we observe that some securities show consistent relationships across different time intervals, while others exhibit more transient interactions. These findings suggest that certain securities are more interconnected in terms of their responses to market movements, while others may be more isolated or reactive to specific market conditions. The heatmap allows us to identify key securities that act as central nodes in the network, with strong connections across multiple market dimensions.

In addition to the intra-layer relationships captured by the heatmap, we also focus on interlayer connections, which represent the influence of one market dimension on another over time. These interlayer relationships model how variables from one layer (e.g., asset prices in Layer 1) can influence variables in another layer (e.g., returns in Layer 2) with a time lag, capturing the spillover effect. To visualize the interlayer connections, we create a separate heatmap that highlights the relationships between the layers themselves. This heatmap focuses specifically on the interlayer edges, where the relationships between nodes across different layers are modeled with a time lag of \( p \) days. The heatmap provides a clear picture of how asset prices, trading values, and returns influence each other over time.
\begin{figure}[ht]
    \centering
    \includegraphics[width=\linewidth]{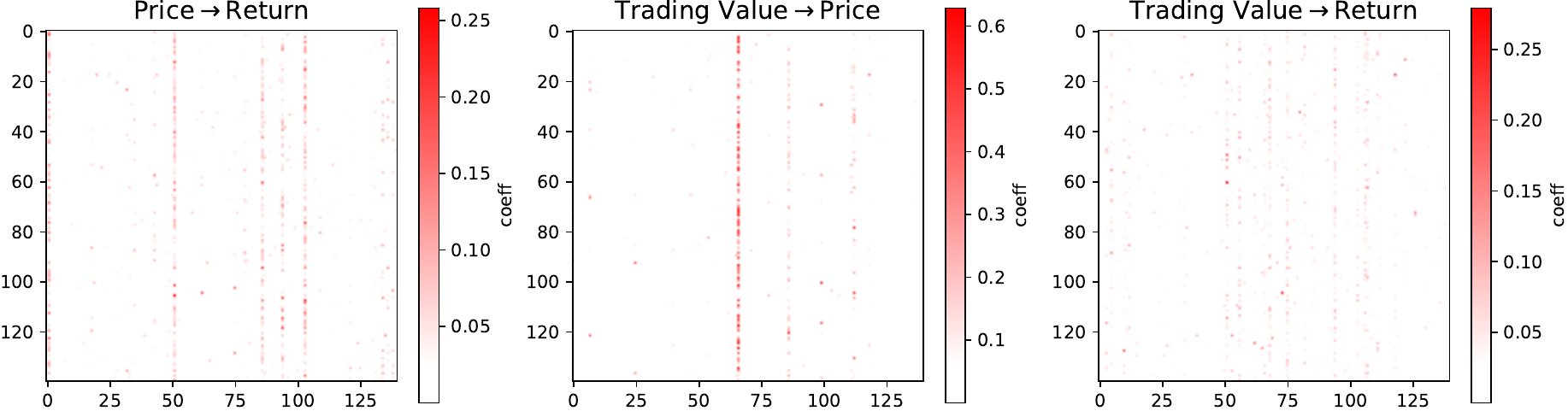} 
    \caption{Heatmaps showing the evolution of interlayer connection strengths at time \( t=95 \) for the relationships: \(\text{Price} \rightarrow \text{Return}\), \(\text{Trading Value} \rightarrow \text{Price}\), and \(\text{Trading Value} \rightarrow \text{Return}\). The color intensity, mapped from white to red, indicates the magnitude of the connection coefficients derived from the Random Forest model, highlighting the directional spillover effects between layers.}
    \label{fig:interlayer}
\end{figure}

Figure~\ref{fig:interlayer} displays the interlayer connections between asset prices, trading values, and returns, highlighting the spillover effects across layers. By examining the heatmap, we can identify which market variables exert the strongest influence on others over time, as well as the temporal delays in the relationship between layers. This visualization provides insights into how changes in one market dimension (e.g., price) propagate and affect other dimensions (e.g., returns), which is crucial for understanding the broader dynamics of the financial market.

\subsection{LSTM Feature Selection}

To identify the most relevant features for predicting financial crises, we evaluated all states of spillover relationships derived from the multilayer network. Among these, two degree-based features—trading value, return interlayer and trading value—exhibited significant positive correlations with crisis prediction targets.

This finding aligns with intuition, as these features capture the intensity of spillover effects from trading values to returns and the overall trading value connectivity, both critical indicators of market stress propagation. The robustness of this relationship was further validated using ridge linear regression analysis, detailed in Appendix~B. The regression results demonstrated a peak coefficient of determination (\(R^2\)) around 0.4 and a correlation coefficient of approximately 0.6, supporting the predictive power of the selected features.

\begin{figure}[ht]
    \centering
    \includegraphics[width=\linewidth]{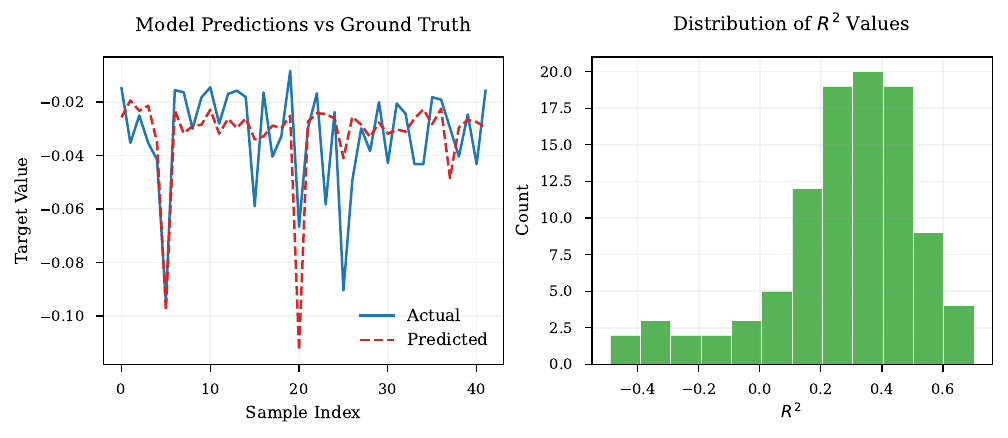} 
    \caption{(Top) Distribution of \(R^2\) values from ridge regression analysis showing a peak around 0.4, indicating strong explanatory power of the selected features. (Bottom) Correlation plot between crisis prediction targets and the features trading value, return interlayer and trading value demonstrating a positive correlation near 0.6. These results support the significance of these features in the LSTM-based crisis prediction model.}
    \label{fig:Ridge_linear_relation}
\end{figure}

Figure~\ref{fig:Ridge_linear_relation} in below illustrate the distribution of \(R^2\) values and the strength of correlation, respectively, confirming that trading value, return interlayer and trading value are key predictors in the LSTM crisis prediction model.
and in addition Appendix~B shows analyze of each relation.

\subsection{LSTM and Prediction Result}

In this section, we evaluate the performance of the Long Short-Term Memory (LSTM) model in forecasting the standardized returns of the S\&P 500 securities based on the multilayer network features derived previously. The model is trained on the time series data of the minimum returns calculated over sliding windows, and its predictive accuracy is assessed by comparing the predicted returns with the actual observed returns.

\begin{figure}[ht]
    \centering
    \includegraphics[width=\textwidth]{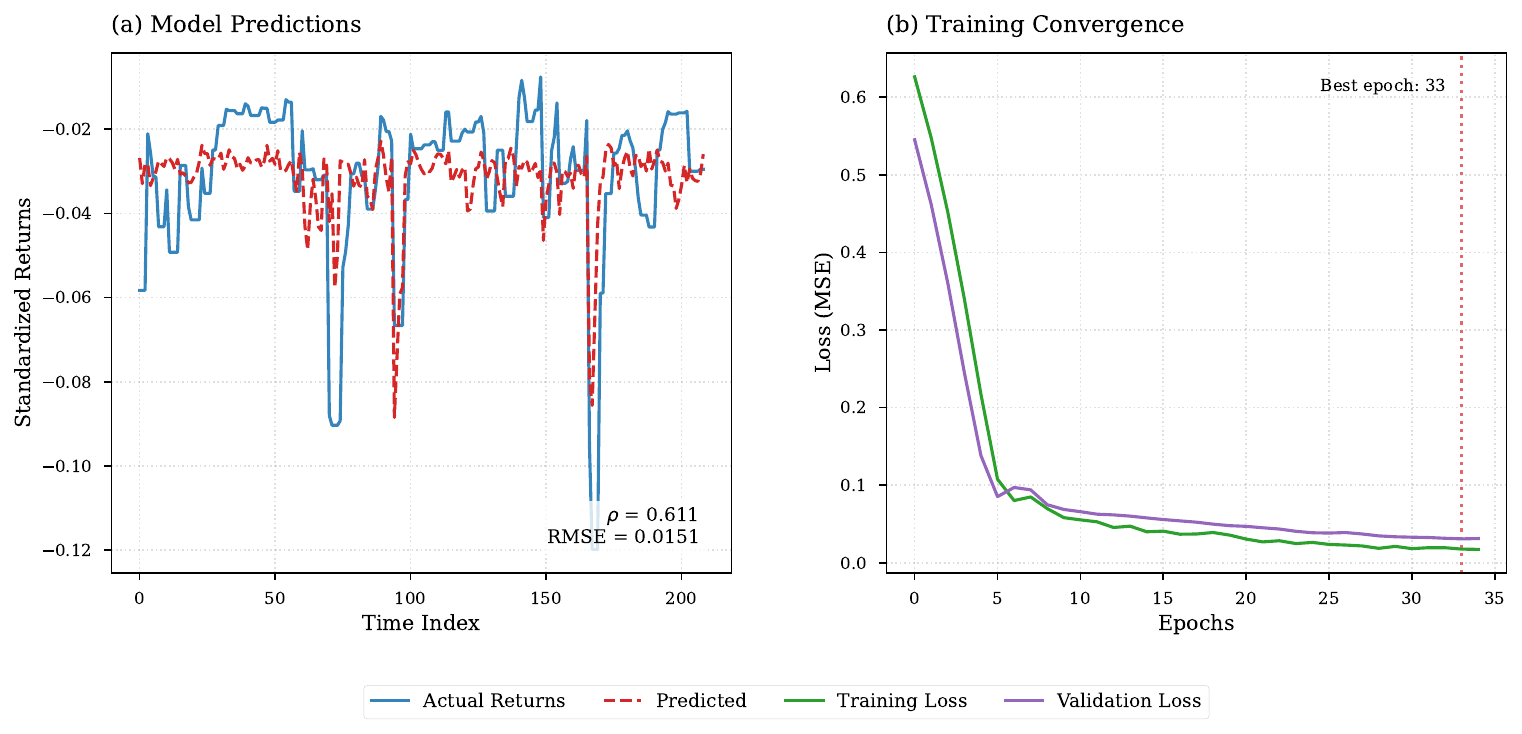} 
    \caption{(a) Actual vs. predicted standardized returns of the S\&P 500 securities over the test period. (b) Training and validation loss (MSE) curves showing model convergence across epochs. The red dashed line indicates the epoch with the best validation loss.}
    \label{fig:lstm_performance}
\end{figure}

Figure~\ref{fig:lstm_performance} presents two key visualizations: (a) the comparison of actual versus predicted returns over the test period, and (b) the training and validation loss curves over the epochs of model training. The plots illustrate the model's forecasting capability and the convergence behavior during training.

The left panel (a) shows the time series of actual returns (solid blue line) alongside the model's predictions (dashed red line). Visual inspection reveals that the LSTM model effectively captures the main trends and fluctuations in returns. The Pearson correlation coefficient \( \rho \) between the actual and predicted returns and the root mean squared error (RMSE) are displayed on the plot, quantifying the model's predictive performance.

The right panel (b) displays the mean squared error loss during training and validation phases. The training converges smoothly with a clear decrease in loss, and the validation loss curve suggests good generalization without significant overfitting. The vertical red dotted line marks the epoch with the lowest validation loss, representing the optimal stopping point.

These results demonstrate the LSTM model's ability to learn temporal dependencies in the complex financial data derived from the multilayer network, providing accurate short-term forecasts of market returns. The model's performance confirms the utility of the multilayer features in enhancing market crisis prediction and systemic risk assessment.

\begin{figure}[ht]
    \centering
    \includegraphics[width=0.5\textwidth]{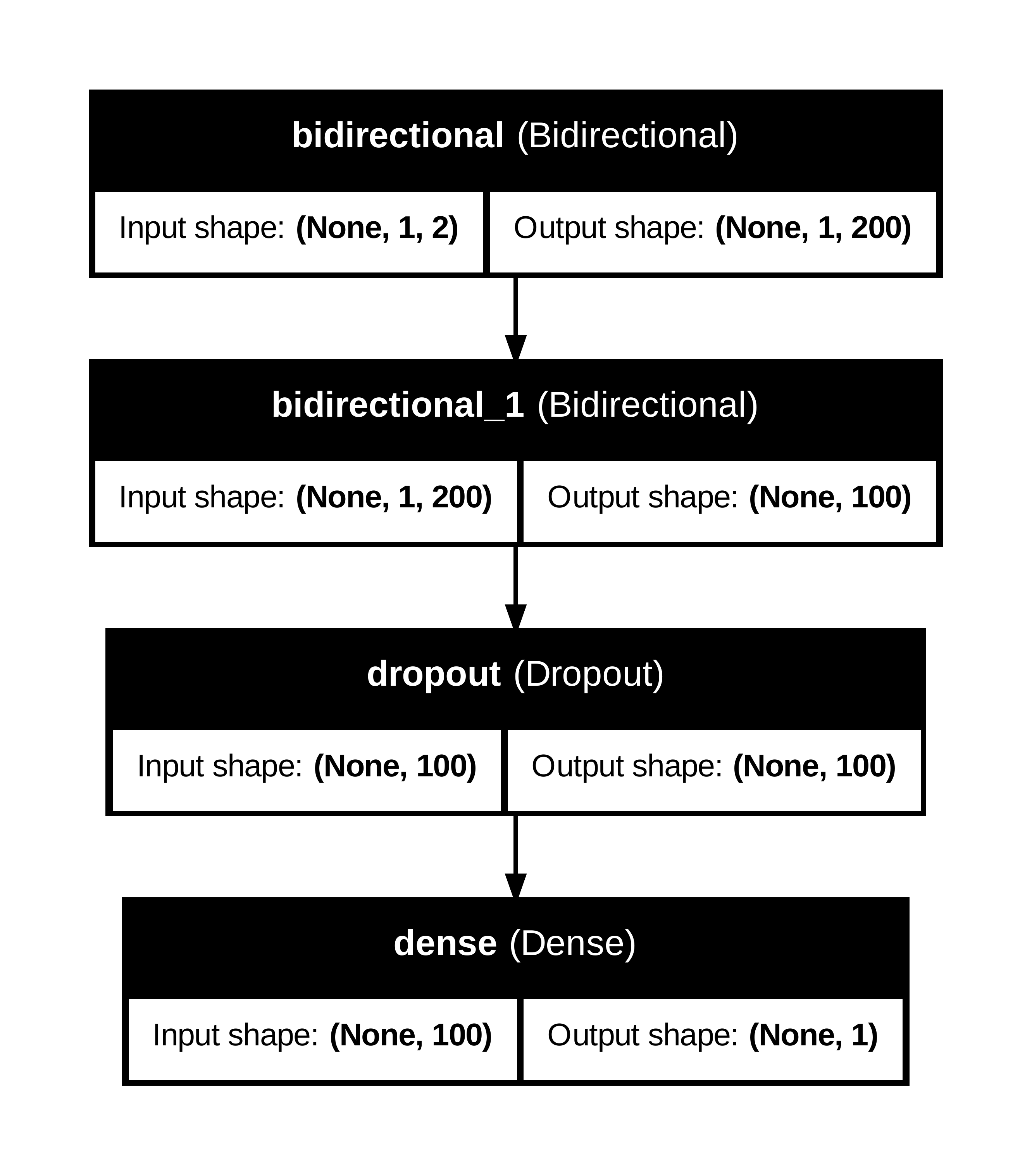} 
    \caption{Architecture of the bidirectional LSTM model used for forecasting standardized financial returns. The model comprises two stacked bidirectional LSTM layers with 100 and 50 units respectively, followed by a dropout layer to reduce overfitting, and a dense output layer producing a single continuous prediction.}
    \label{fig:lstm_architecture}
\end{figure}

The architecture of the LSTM model is designed to capture temporal dependencies in the financial time series data for return prediction. The model consists of two stacked Bidirectional LSTM layers, where the first layer has 100 units and returns sequences to allow the second layer, with 50 units, to further learn complex patterns from the entire sequence. A dropout layer with a rate of 0.3 is added after the second LSTM layer to prevent overfitting by randomly deactivating neurons during training. Finally, a dense layer with a single neuron produces the output prediction. The model is compiled using the Adam optimizer with a learning rate of 0.001 and trained to minimize the mean squared error loss.

\clearpage
\section{Conclusion}

This study presents a novel multilayer network framework integrated with advanced machine learning techniques to improve financial crisis prediction. By modeling financial markets as interconnected layers representing price, return, and trading value, and capturing both within-layer and cross-layer dependencies using Granger causality and Random Forest, we uncover meaningful spillover effects that traditional single-layer approaches may overlook. Our approach leverages Long Short-Term Memory (LSTM) networks to effectively learn temporal patterns from these multilayer features, demonstrating improved forecasting accuracy for market instability.

The identification of key degree-based features, particularly those reflecting interlayer spillovers from trading value to returns, highlights the critical role of cross-market dynamics in systemic risk propagation. The positive correlations and robust predictive performance validated through ridge regression emphasize the model's practical utility. Overall, this integrated methodology offers a comprehensive and interpretable tool for monitoring systemic risk and enhancing early warning capabilities in complex financial systems.

Future work may explore extending this framework to include additional market layers, incorporate alternative causality measures, and evaluate real-time crisis detection performance. By advancing the understanding of multilayer financial networks and harnessing machine learning, this research contributes toward more resilient and responsive financial risk management.


\appendix
\clearpage
\section{Security ID Mapping}
All security IDs used throughout the graphs and analyses correspond to the tickers listed in Table~\ref{tab:security_ids}.
\begin{table}[htbp]
\centering
\caption{Mapping of Security IDs to Ticker Symbols}
\label{tab:security_ids}
\begin{tabular}{rl|rl|rl|rl}
\hline
ID & Ticker & ID & Ticker & ID & Ticker & ID & Ticker \\
\hline
0 & AAPL & 35 & DE   & 70 & LNC  & 105 & ROST \\
1 & ABT  & 36 & DHI  & 71 & LOW  & 106 & RSG  \\
2 & ADBE & 37 & DHR  & 72 & LRCX & 107 & RTX  \\
3 & ADP  & 38 & DIS  & 73 & LUV  & 108 & SBUX \\
4 & ADSK & 39 & DUK  & 74 & MCD  & 109 & SHW  \\
5 & AEP  & 40 & DXC  & 75 & MCK  & 110 & SLB  \\
6 & AFL  & 41 & EL   & 76 & MCO  & 111 & SPG  \\
7 & AIG  & 42 & EOG  & 77 & MDT  & 112 & SPGI \\
8 & ALL  & 43 & ES   & 78 & MMC  & 113 & STT  \\
9 & AMGN & 44 & ETN  & 79 & MMM  & 114 & STZ  \\
10 & AMT & 45 & EXC  & 80 & MRK  & 115 & SWK  \\
11 & AMZN & 46 & EXPD & 81 & MS   & 116 & SWKS \\
12 & AON  & 47 & F    & 82 & MSFT & 117 & SYK  \\
13 & APD  & 48 & FCX  & 83 & MTD  & 118 & SYY  \\
14 & AXP  & 49 & GD   & 84 & MU   & 119 & T    \\
15 & BA   & 50 & GE   & 85 & NEE  & 120 & TAP  \\
16 & BAX  & 51 & GILD & 86 & NEM  & 121 & TGT  \\
17 & BIIB & 52 & GS   & 87 & NKE  & 122 & TMO  \\
18 & BKNG & 53 & HD   & 88 & NOC  & 123 & TROW \\
19 & BKR  & 54 & HIG  & 89 & NSC  & 124 & TRV  \\
20 & BMY  & 55 & HON  & 90 & NUE  & 125 & TSCO \\
21 & C    & 56 & HUM  & 91 & NVDA & 126 & TXN  \\
22 & CAT  & 57 & IDXX & 92 & O    & 127 & UNH  \\
23 & CHD  & 58 & INTC & 93 & ORCL & 128 & UNP  \\
24 & CHKP & 59 & INTU & 94 & OXY  & 129 & VFC  \\
25 & CI   & 60 & IT   & 95 & PEP  & 130 & VLO  \\
26 & COF  & 61 & ITW  & 96 & PFE  & 131 & VMC  \\
27 & COP  & 62 & JNJ  & 97 & PG   & 132 & VRTX \\
28 & COST & 63 & JPM  & 98 & PGR  & 133 & VZ   \\
29 & CSCO & 64 & KLAC & 99 & PLD  & 134 & WBA  \\
30 & CSX  & 65 & KMB  & 100 & PLUG & 135 & WDC  \\
31 & CTAS & 66 & KO   & 101 & PNC  & 136 & WEC  \\
32 & CTSH & 67 & LHX  & 102 & QCOM & 137 & WFC  \\
33 & CVS  & 68 & LLY  & 103 & REGN & 138 & WMT  \\
34 & CVX  & 69 & LMT  & 104 & RMD  & 139 & XOM  \\
\hline
\end{tabular}
\end{table}

\clearpage
\section{Feature Selection Analysis}
\label{appendix:feature_selection}

\begin{figure}[ht]
    \centering
    \includegraphics[width=0.9\textwidth]{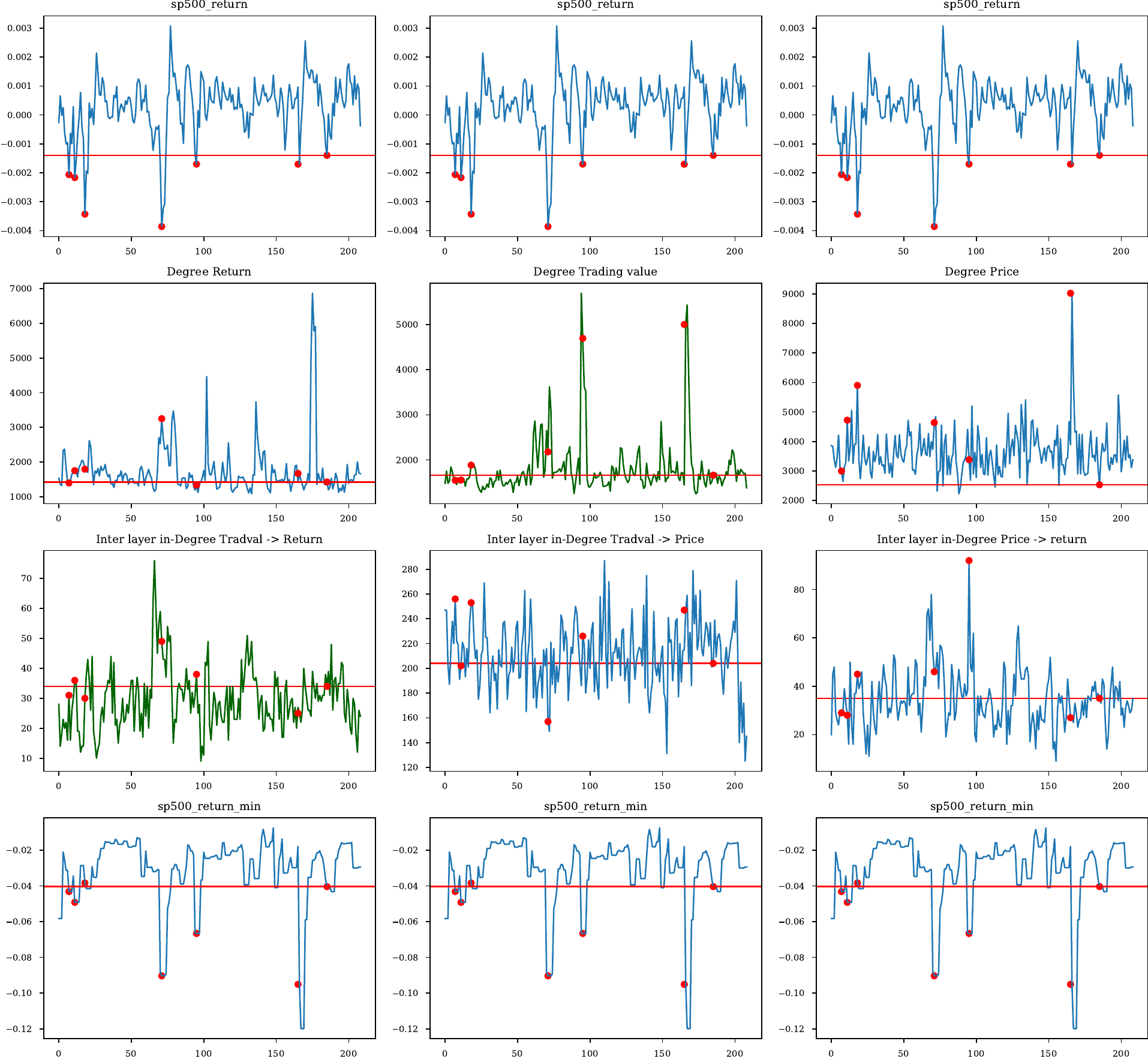}
    \caption{The green curves represent features exhibiting stronger correlation with crisis periods, while the red points mark crisis events occurring within the data timeframe. This visualization highlights the relationship between selected degree-based features and financial crises.}
    \label{fig:ridge_r2_correlation}
\end{figure}

\vspace{1em}

\begin{table}[htbp]
\centering
\caption{Correlation coefficients (\%) between degree features and various return measures. Negative values indicate negative correlations and positive values indicate positive correlations.}
\label{tab:degree_correlations}
\begin{tabular}{l r}
\hline
\multicolumn{2}{c}{\textbf{Degree interlayer trading value \(\rightarrow\) return, return}} \\
\hline
Correlation with mean return & -27.34\% \\
Correlation with minimum return* & -19.97\% \\
Correlation with maximum return & 27.73\% \\
Correlation with return variance & 26.11\% \\
\hline
\multicolumn{2}{c}{\textbf{Degree trading value, return}} \\
\hline
Correlation with mean return & -35.56\% \\
Correlation with minimum return* & -62.55\% \\
Correlation with maximum return & 46.65\% \\
Correlation with return variance & 48.79\% \\
\hline
\end{tabular}
\end{table}

\medskip

\clearpage
\bibliography{reference}

\end{document}